\documentclass[aps,pre,superscriptaddress,preprint]{revtex4}
\PassOptionsToPackage{sort&compress}{natbib}
\usepackage{graphicx}
\usepackage{float}
\usepackage{amsmath,amssymb,amsfonts}
\usepackage{array}
\usepackage{xcolor}
\usepackage{tabularx}

\usepackage[caption=false]{subfig}

\newcommand{\intR}{\int_{\mathbb R}}
\newcommand{\intS}{\int_{\mathbb S}}

\begin{document}
	
\title{Synchronization of coupled second-order Kuramoto-Sakaguchi oscillators}

\author{Jian Gao}
\affiliation{Bernoulli Institute for Mathematics, Computer Science, and Artificial Intelligence, University of Groningen, P.O. Box 407, 9700 AK, Groningen, The Netherlands}
	
\author{Konstantinos Efstathiou}\email{k.efstathiou@dukekunshan.edu.cn}
\affiliation{Division of Natural and Applied Sciences and Zu Chongzhi Center for Mathematics and Computational Science, Duke Kunshan University, No. 8 Duke Avenue, Kunshan 215316, China}
	
\begin{abstract}
We study the synchronization of oscillators with inertias and phase shifts, namely the second-order Kuramoto-Sakaguchi model. Using the self-consistent method, we find that the effect of inertia is the introduction of effective phase shifts. The discontinuous synchronization transition of the Kuramoto-Sakaguchi model changes to a continuous one when the value of inertia is small. In addition, we find a new synchronization process, in which with increasing coupling strength the system reaches an oscillating state instead of complete synchronization due to the cross-effect of phase shifts and inertias. Through numerical simulations, the same type of synchronization process is also found for oscillators in complex networks, including scale-free, small-world and random networks. 
\end{abstract}

\maketitle

\section{Introduction}

Synchronization of oscillators, especially synchronization in complex networks \cite{Arenas2008}, has been recognized as one of the important phenomena in nature. Among the different models of oscillator dynamics, the Kuramoto model \cite{Kuramoto1987}, and its various generalizations \cite{Rodrigues2016}, are some of the most popular models. Within this class of generalized Kuramoto models, second-order oscillator models, that is, oscillators with inertias, have been used for describing the dynamics of fireflies \cite{Ermentrout1991}, Josephson junction arrays \cite{Levi1978, Watanabe1994, Trees2005}, goods markets \cite{Ikeda2012}, dendritic neurons \cite{Sakyte2011}, and power grids \cite{Filatrella2008}. Due to the effect of inertias, phenomena such as hysteresis, bi-stability and abrupt transitions are found for these second-order oscillators \cite{Tanaka1997, Tanaka1997a, Gao2017}. In \cite{Tanaka1997, Gao2017} the changes from continuous to abrupt phase transition for second-order oscillators have been studied in detail using the self-consistent method.

As a natural generalization, oscillators with both inertias and phase shifts, namely the second-order Kuramoto-Sakaguchi model, are considered in \cite{Barre2016}. It has been found that due to the effect of inertias the synchronization transition of oscillators can be changed from continuous to abrupt and vice versa.  In this paper, we generalize the self-consistent method presented in \cite{Gao2017} to the second-order Kuramoto-Sakaguchi model. We find that the inertias introduce effective phase shifts and that the type of synchronization transition is affected by the mixture of these inertia-induced phase shifts and the ones built into the model. 

Moreover, we find a new type of synchronization process with increasing coupling strength. In this process, oscillators converge to an oscillating state by forming several synchronization clusters, which cannot be further synchronized by increasing the coupling strength. This process is quite different from the common belief that with sufficient large coupling strengths the coupled Kuramoto-like oscillators are typically synchronized to a highly coherent steady state, except for some specific choice of parameters, such as with phase shift $\pm\pi/2$. Through the self-consistent method and dynamical analysis of the synchronized clusters, we show that this process is due to the cross-effect of inertias and phase shifts, and is not limited to the case of all-connected oscillators. Through numerical simulations, this new type of synchronization process is also found in oscillators connected in complex networks.

Our paper is organized as follows. In Section \ref{section_one}, we generalize the self-consistent method to oscillators with inertias and phase shifts. The mixture of effective (inertia-induced) and intrinsic phase shifts is associated to the change of properties of the synchronization transition. Using the self-consistent method, in Section \ref{section_two}, we find the new synchronization process to oscillating states and study it through the self-consistent method and dynamical analysis. Using numerical simulations, this process is also observed for oscillators on complex networks. We conclude this paper in Section \ref{section_four}.

\section{Effective Phase Shifts}\label{section_one}

To focus on the effect of phase shifts, we assume that all the oscillators have the same inertia $m$ and damping constant $D$. The dynamics of the second-order Kuramoto-Sakaguchi model reads 
\begin{equation}\label{eq_dynamics_two}
m\ddot{\varphi}_i+D\dot{\varphi}_i=\Omega_i+\frac{K}{N}\sum_{j=1}^{N}\sin(\varphi_j-\varphi_i-\alpha), \ i=1,2\dots,N
\end{equation}
where $N$ is the number of oscillators and $K$ is the uniform coupling strength. Each oscillator is described by its phase $\varphi_i\in\mathbb{S}$ with $\Omega_i$ as its natural frequency. The intrinsic phase shift $\alpha$ is added in the coupling term $\sin(\varphi_j-\varphi_i-\alpha)$. The standard second-order model corresponds to $\alpha = 0$.

Following the work by Kuramoto \cite{Kuramoto1987} we define the order parameter
\begin{equation}\label{eq_order_definition}
re^{i\phi}=\frac{1}{N}\sum_{j=1}^{N}e^{i\varphi_j},
\end{equation}
where $r$ and $\phi$ represent the coherence and mean-phase of the oscillators. If all the oscillators run independently, their phases will almost uniformly distribute along the unit circle. As a result, we have $r\approxeq 0$, and the state is called incoherence. On the other hand, if all of the oscillators are synchronized and have the same phase $\theta_i(t)\equiv\theta(t)$, we have $r=1$. This is called the complete synchronization state of the system. 

Using $r$ and $\phi$, the model \eqref{eq_dynamics_two} can also be rewritten in a mean-field form as
\begin{equation}\label{eq_dynamics_three}
m\ddot{\varphi}+D\dot{\varphi}=\Omega + Kr(t)\sin(\phi(t)-\varphi-\alpha),
\end{equation}
where the subscripts have been dropped. In Eq.~\eqref{eq_dynamics_three} each oscillator interacts with other oscillators only through the mean-field terms $r$ and $\phi$. Therefore, the dynamics of the system can be obtained through the analysis of each single oscillator with a presupposed mean-field. 

For simplicity, in this paper we assume an infinite number of oscillators $N\rightarrow\infty$, and that the distribution of natural frequencies of oscillators $g_\Omega(\Omega)$ is symmetric, $g_\Omega(\Omega)=g_\Omega(-\Omega)$, and unimodal. The essential states of the system are the steady states defined as
\begin{equation}\label{def_steady_eq}
r(t) = r, \ \ \phi(t)=\Omega^r t +\Psi,
\end{equation}
where the order parameter $r(t)$ is independent of time, and the phase $\phi(t)$ has a constant rotation velocity. Without loss of generality, we set $\Psi\equiv0$. Define the phases $\theta$ of each oscillator in a rotating coordinate frame through the transformation $\theta = \varphi -\phi(t)$. Substitution of Eq.~\eqref{def_steady_eq} into Eq.~\eqref{eq_dynamics_three} yields 
\begin{equation}\label{eq_phase_differece}
m\ddot{\theta}+D\dot{\theta}=(\Omega-D\Omega^r)-Kr\sin(\theta+\alpha).\ \ 
\end{equation}
For $\alpha=0$, Eq.~\eqref{eq_phase_differece} is exactly the same as the one for a single second-order oscillator without intrinsic phase shift \cite{Gao2017}. Following \cite{Gao2017},  Eq.~\eqref{eq_phase_differece} can be rewritten in the standard form as 
\begin{equation}\label{eq_single_dynamic}
\ddot{\theta}+a\dot{\theta}=b-\sin(\theta+\alpha),
\end{equation}
with rescaled time $\tau=t/\sqrt{m/Kr}$ and
\begin{equation}\label{def_transformation}
a=\frac{D}{\sqrt{Krm} }, \ \  b=\frac{\Omega-D\Omega^r}{Kr}.
\end{equation}
Because of its dependence on $\Omega$, the parameter $b$ follows the distribution $g_b(b)=Krg_\Omega(Krb+D\Omega^r)$.

It is known from the earlier studies \cite{Tanaka1997,Tanaka1997a,Gao2017} that the system Eq.~\eqref{eq_single_dynamic} has two stable states, one fixed point and one limit cycle \cite{Strogatz2014,Tanaka1997a}. The rotation frequency of oscillators is defined as $\omega=\dot{\theta}$. Taking $a>0$, the stable fixed point reads
\begin{equation}
\label{estimation of fixed point}
\theta_0=\arcsin(b)-\alpha,\ \  \omega_0=0,
\end{equation}
with the existence condition $b\leq b_L(a)=1$. On the other hand, for the limit cycle, using the same estimation method as in \cite{Gao2017} we have the approximate expression $\dot\theta(\tau)$ for the limit cycle, given by
\begin{equation}\label{limit cycle}
\dot\theta(\tau)
= \frac{b}{a} - \frac{1}{a}\sigma \sin(\theta(\tau)+\Delta+\alpha),
\end{equation}
where the coupling factor $\sigma$ and phase shift term $\Delta$ are
\begin{equation}\label{def_transformation2a}
\sigma=\frac{a^2}{\sqrt{b^2+a^4}},\ \ \Delta=\arcsin \left(\frac{-b}{\sqrt{b^2+a^4}} \right).
\end{equation}
The existence condition of the limit cycle can be calculated through Melnikov's method \cite{Guckenheimer2013} or Lyapunov’s direct method \cite{Risken1996fokker} and numerical simulations \cite{Gao2017} as
\begin{equation}\label{bs}
b\geq b_S=
\begin{cases}
(4/\pi)\,a-0.305a^3, \ \ &a\leq 1.193,\\
1, \ \ &a>1.193.
\end{cases}
\end{equation}
Eq.~\eqref{limit cycle} shows that running oscillators have the same dynamics as Kuramoto-Sakaguchi oscillators with coupling factor $\sigma$ and \emph{effective phase shift} $\alpha+\Delta$ as the combination of intrinsic phase shift $\alpha$ and inertia-induced phase shift $\Delta\in(-\pi/2,\pi/2)$.

\begin{figure}
	\centering
	\includegraphics[width=0.9\textwidth]{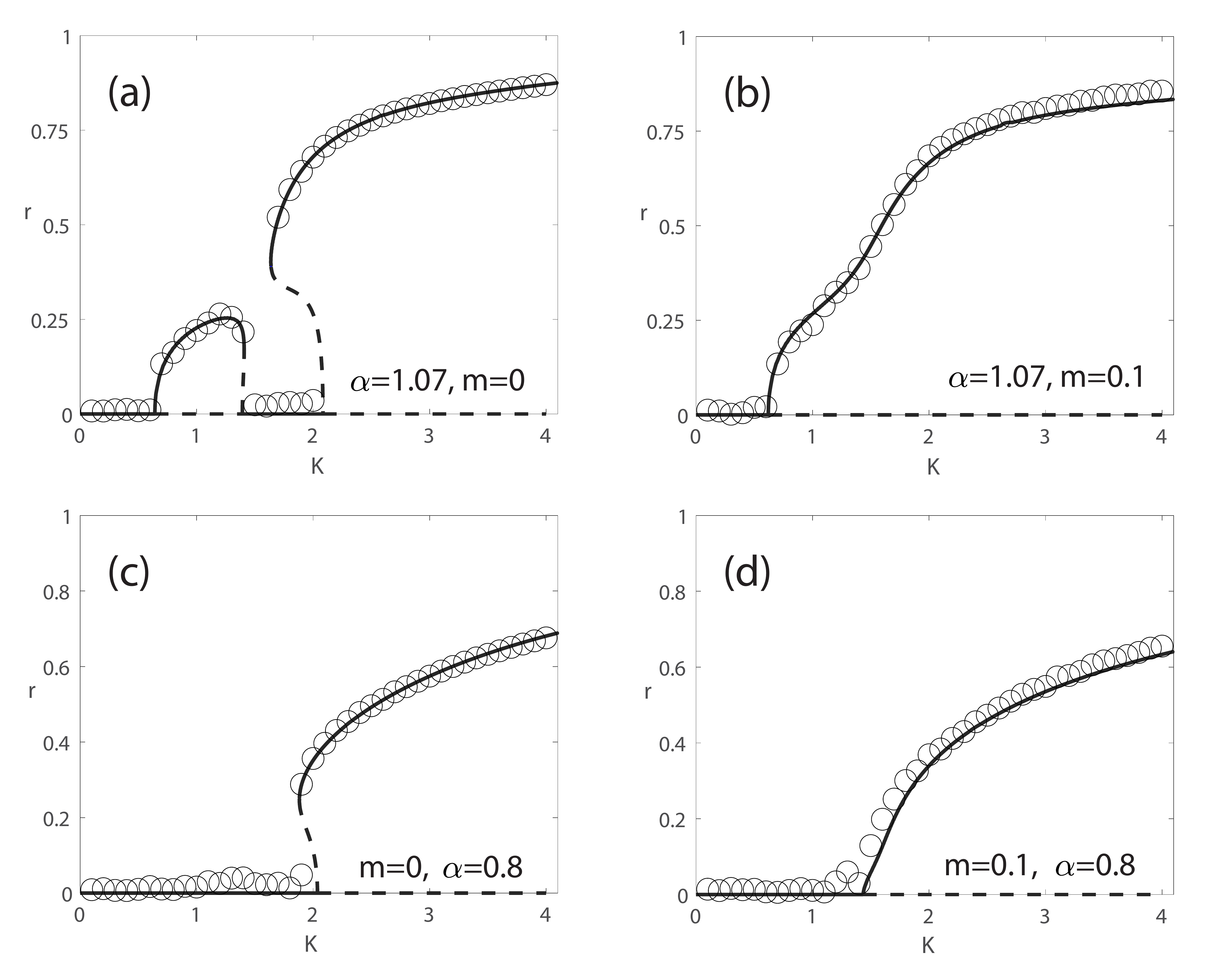}
	\caption{Synchronization  for oscillators with a double Gaussian distribution $g_\Omega(\Omega)=0.6\times\frac{1}{\sqrt{2\pi}}e^{-\Omega^2/2}+0.4\times\frac{1}{\sqrt{2\pi}\times0.1}e^{-\Omega^2/(2\times0.1^2)}$ with $\alpha=1.07$ and different inertias $m=0$ (a), $m=0.1$ (b);  or a double Lorentz distribution $g_\Omega(\Omega)=0.8\times\frac{1}{\pi}\frac{1}{\Omega^2+1}+0.2\times\frac{1}{\pi}\frac{0.075}{\Omega^2+0.075^2}$ with $\alpha=0.8$ and different inertias, $m=0$ (c), $m=0.1$(d). The solid (dashed) lines are solutions of self-consistent equations, and correspond to stable (unstable) steady states. Circles are from the numerical simulations of $10000$ oscillators with proper initial states.}\label{fig_three}
\end{figure}

Even a small inertia value can introduce the mixture effect of phase shifts $\alpha$ and $\Delta$. As a result, several non-trivial transitions of Kuramoto-Sakaguchi oscillators that depend on the specific choice of phase shifts will be undermined by inertias. These include the non-universal transition processes in \cite{Omel2012,Omel2013}, shown in Fig.~\ref{fig_three}(a-b), and the discontinuous transition Fig.~\ref{fig_three}(c-d). Note that with the introduction of inertias, the transition processes are not always changed from continuous to abrupt. The opposite also happens when there are phase shifts, as pointed out in \cite{Barre2016} using the stability analysis around the critical point.

When the inertia is not zero, there is a bistable parameter region, where the system has both a fixed point and a limit cycle, given by $b_L\geq b \geq b_S$. Each oscillator is either \emph{locked} at the fixed point or \emph{running} along the limit cycle. Taking $N\rightarrow\infty$, the order parameter defined in Eq.~\eqref{eq_order_definition} can be rewritten as
\begin{equation}\label{Eq.model.3}
r=\intR \intS \intR e^{i\theta(t)} p(\Omega,\theta_0,\omega_0) d\omega_0 d\theta_0 d\Omega,
\end{equation}
where $p(\Omega,\theta_0,\omega_0)$ represents the distribution of initial conditions and natural frequencies, and the dynamics $\theta(t)$ for each oscillator depends on its initial conditions and $\Omega$.
Note that $\intS\intR p(\Omega,\theta_0,\omega_0) d\theta_0 d\omega_0 = g_\Omega(\Omega)$.
If we know the ratio of locked and running oscillators in the system then the last expression can be simplified.
Substituting the solution of locked and running oscillators, Eq.~\eqref{estimation of fixed point} and Eq.~\eqref{limit cycle}, into Eq.~\eqref{Eq.model.3}, together with their existence conditions, we have the self-consistent equations 
\begin{subequations}\label{eq_self_consistent}
\begin{align}
	r&=\intR  g_\Omega(\Omega)\rho_{l}(a,b)\left(\sqrt{1-b^2}\cos\alpha-b\sin\alpha\right) \notag \\
	&-g_\Omega(\Omega)\rho_{r}(a,b)\left(\frac{b}{\sigma}+\sqrt{\frac{b^2}{\sigma^2}-1}\right)\sin(\Delta+\alpha)d\Omega,\\
	0&=\intR  g_\Omega(\Omega)\rho_{l}(a,b)\left(b\cos\alpha-\sqrt{1-b^2}\sin\alpha\right) \notag \\
	&+g_\Omega(\Omega)\rho_{r}(a,b)\left(\frac{b}{\sigma}-\sqrt{\frac{b^2}{\sigma^2}-1}\right)\cos(\Delta+\alpha)d\Omega.
\end{align}
\end{subequations}
The fraction functions $\rho_{l}(a,b)$ and $\rho_{r}(a,b)$ are the fraction of locked and running oscillators respectively, satisfying the normalization condition $\rho_{l}(a,b)+\rho_{r}(a,b)=1$,
and the boundaries $\mathbf{1}_S(a,b)\leq\rho_{l}(a,b)\leq\mathbf{1}_L(a,b)$.
The indicator functions $\mathbf{1}_{S,L}$ take the value $1$ if $|b| < b_S(a)$ or $|b| < b_L(a)$ and $0$ otherwise, corresponding to cases of running or locked oscillators. The most commonly used fraction functions are the two indicator functions $\rho_{l}(a,b)=\mathbf{1}_S(a,b)$ and $\rho_{l}(a,b)=\mathbf{1}_L(a,b)$ where all the oscillators are in the limit cycle state, or the fixed point state, as long as it is possible. These two functions correspond to the so-called forward and backward processes. In the forward process, the initial state for small coupling strength is the incoherence state, and the coupling strength is then progressively increased. In the backward process, the initial state for large coupling strength is the synchronization state, and the coupling strength is then progressively decreased. For second-order oscillators, these two processes in general do not coincide with each other, a phenomenon known as \emph{hysteresis} \cite{Tanaka1997,Gao2017}.

Compared with the previous results, it is easy to verify that when $\alpha=0$, from Eq.~\eqref{eq_self_consistent} one regains the self-consistent equations for second-order oscillators without phase shifts in \cite{Gao2017} using the approximation $b/\sigma-\sqrt{b^2/\sigma^2-1}\approx \sigma/(2b)$ which is valid for small $\sigma$. On the other hand, in the limit $m\rightarrow0$, we have $b_{S,L}(a)\rightarrow1, \sigma\rightarrow1, \Delta\rightarrow 0$. The self-consistent equations \eqref{eq_self_consistent} in this case are the same as the ones obtained for Kuramoto-Sakaguchi models in \cite{Omel2012,Omel2013}, 

Following \cite{Omel2013,Gao2017}, by defining $q=Kr$ and correspondingly $a=D/\sqrt{qm}$ and $g_b(b)=qg_\Omega(qb+D\Omega^r)$, the self-consistent equations \eqref{eq_self_consistent} can be rewritten as
\begin{subequations}\label{eq_self_consistent3}
	\begin{align}
	&\begin{aligned}
    \frac{\cos\alpha}{K}=&F_1(q,\Omega^r)\equiv
	\int_{-\infty}^{\infty}g_\Omega(qb+D\Omega^r)\\
	&\left[\rho_l\sqrt{1-b^2}+\rho_r\left(\frac{b}{\sigma}-\sqrt{\frac{b^2}{\sigma^2}-1}\right)\sin\Delta\right]db,
	\end{aligned}\\
	&\begin{aligned}
	\frac{\sin\alpha}{K}=&F_2(q,\Omega^r)\equiv
	\int_{-\infty}^{\infty}g_\Omega(qb+D\Omega^r)\\
	&\left[\rho_lb+\rho_r\left(\frac{b}{\sigma}-\sqrt{\frac{b^2}{\sigma^2}-1}\right)\cos\Delta\right]db,
	\end{aligned}\label{eq_self_consistent3b}
	\end{align}.
\end{subequations}
Eq.~\eqref{eq_self_consistent3} defines a map from $(q,\Omega^r)$ to $(\alpha, K)$. The solutions of the self-consistent equations can be denoted as the quad $(q,\Omega^r,K,\alpha)$ corresponding to points on the graph of this map. From the quad $(q,\Omega^r,K,\alpha)$, it is straightforward to obtain the solutions for the order parameter as the triplets $(K,\alpha,r)$ and $(K,\alpha,\Omega^r)$. These are depicted in Fig.~\ref{fig_three}.

The results of the numerical simulation that demonstrates the mixture effect of intrinsic and inertia-induced phase shifts are shown in Fig.~\ref{fig_three}. Here, we consider $N=10000$ oscillators with either no or small inertias $m=0.1$. The natural frequencies are chosen from a double Gaussian or a double Lorenz distribution considered in \cite{Omel2012,Omel2013}. The coupling strength is increased from $K=0$ to $K=4$ with increment $dK=0.1$. To obtain the stable states at each coupling strength $K$,  two initial states of oscillators are considered. One is the incoherence state, and the other is the synchronization state. From these two initial states, after sufficient long transient time $t=100$, we obtain the stable states at each coupling strength $K$, shown as circles in Fig.~\ref{fig_three}. The theoretical results are obtained from the self-consistent method in \cite{Omel2012,Omel2013} for $m=0$ and the equations Eq.~\eqref{eq_self_consistent3} for $m=0.1$. Due to the fact that the inertia $m=0.1$ is quite small, the difference between $b_S$ and $b_L$ is negligible. The synchronization transitions can be obtained directly from the stable states. If there is only one stable state for each $K$, the transition is continuous. On the contrary, if there are multiple and discontinuous branches of stable states, the transitions are abrupt.

Comparing the numerical simulations and theoretical results, we firstly find that the theoretical predictions of the self-consistent method coincide well with the results of the numerical simulations. Secondly, even with a small value of inertias, such as $m=0.1$, the stable states of oscillators change dramatically, resulting in corresponding changes in the synchronization transitions. This phenomenon is found in \cite{Barre2016} through the stability analysis around the critical point. In this paper, through Eq.~\eqref{limit cycle} and the self-consistent method, we show that the physical mechanism of these transitions is the inertia-induced phase shift $\Delta$ and its direct mixture with the intrinsic phase shift $\alpha$. This mixture results in the cancellation of the effect of the phase shift and consequently leads to the continuous synchronization transitions for oscillators with unimodal distributions. Interestingly, this analysis can also be applied to the second-order oscillators with $\alpha=0$, where the phase shift $\Delta$ in general introduces abrupt transitions \cite{Gao2018}.

\section{Oscillating Synchronization Process}\label{section_two}

\begin{figure}
	\centering
	\includegraphics[width=0.9\textwidth]{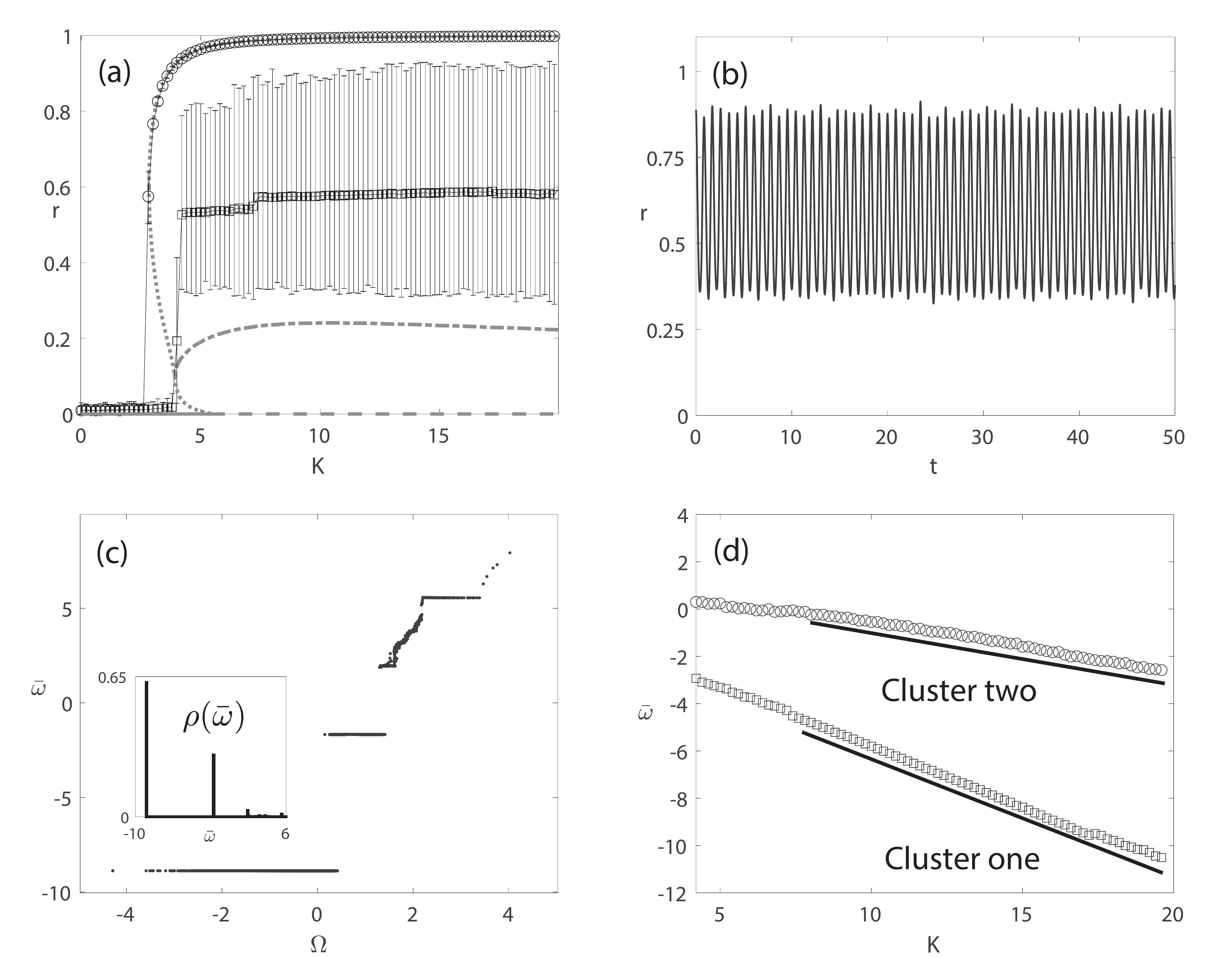}
	\caption{(a) Synchronization transitions for oscillators with a Gaussian distribution $g_\Omega(\Omega)=\frac{1}{\sqrt{2\pi}}e^{-\Omega^2/2}$  with $m=2, \alpha=0.5$ in shown. The dotted and dash-doted lines are the solutions of self-consistent equations in the backward, forward processes. Squares and circles with error bar are from the numerical simulations of $10000$ oscillators in the forward and backward processes, where the error bar is the standard deviation of $r(t)$. At $k=16$ in the forward process, the oscillating state is in (b) with the order parameter $r(t)$, (c) the mean frequencies of oscillators versus their natural frequencies with the distribution of the mean frequencies in the inner figure. (d) The mean frequencies of two largest synchronized clusters in the forward process.}
\label{Fig_oscillating}
\end{figure}

\begin{figure}
	\centering
	\includegraphics[width=0.9\textwidth]{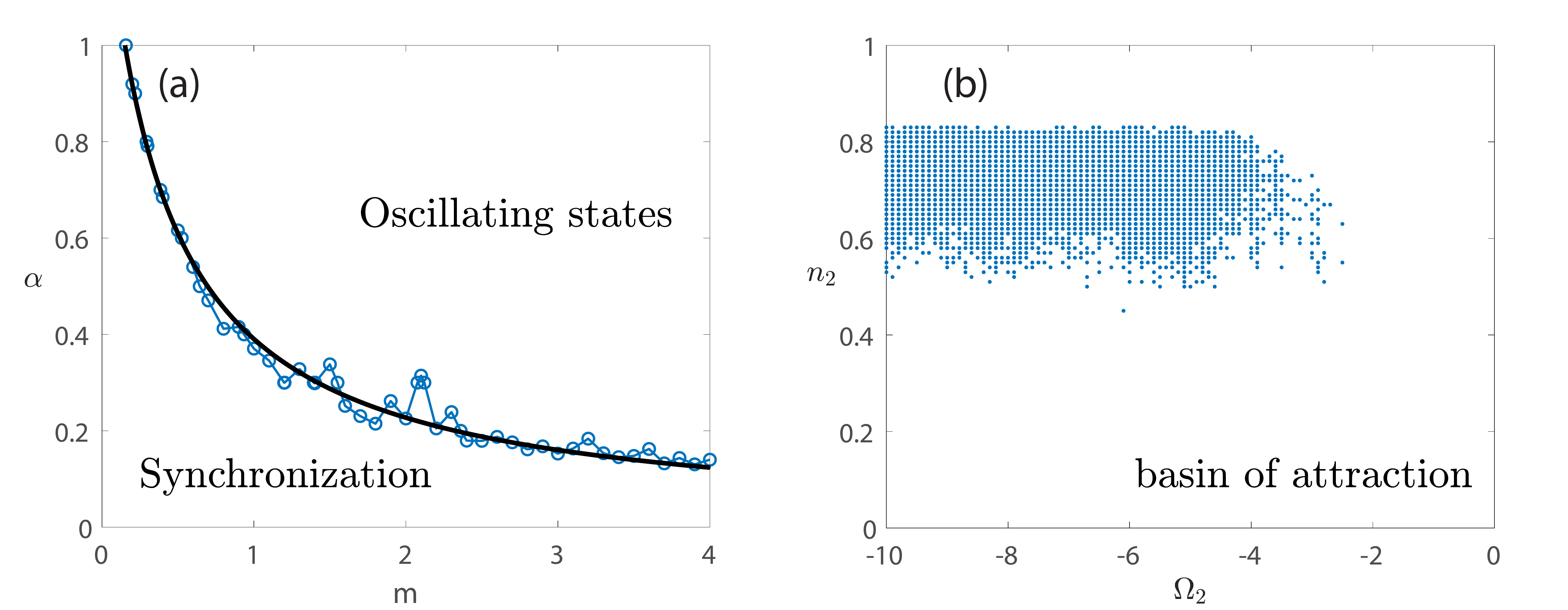}
	\caption{(a) Phase diagram of $1000$ oscillators in the forward processes up to $k=40$ with inertias $m$ and phase shifts $\alpha$. The natural frequencies of oscillators are chosen randomly from a Gaussian distribution $g_\Omega(\Omega)=\frac{1}{\sqrt{2\pi}}e^{-\Omega^2/2}$. (b) The basin of attraction of oscillating states at $K=20$ with $m=2, \alpha=0.5$. The oscillators are sorted separated into two groups by their natural frequencies, with fractions $n_1$ and $n_2$ where $n_1+n_2=1$. The oscillators initial frequencies are chosen randomly from $[\Omega_1-\delta\omega,\Omega_1+\delta\omega]$ and  $[\Omega_2-\delta\omega,\Omega_2+\delta\omega]$ respectively. The initial phases of all oscillators are chosen randomly from $[0,2\pi]$. We set $\Omega_1=-1$ and $\delta\omega=0.1$. The separation of oscillating sates and synchronization states is determined by the standard deviation $\sigma_r=0.1$ of $r(t)$.}\label{Fig_oscillating00 }
\end{figure}

In the previous section we saw how the cross-effect of phase shifts and inertias leads to changes in the synchronization transitions from abrupt to continuous or vice verse through the direct mixture of $\alpha$ and $\Delta$. Here, we show that the same cross-effect to a different synchronization transition where oscillators do not reach a steady state with increasing coupling strength but instead they reach an oscillating state for arbitrarily large coupling strength. This phenomenon is due to the formation of several synchronized clusters and appears in the parameter region of relatively large inertias and phase shifts.

First, we check the existence of complete synchronization state in the limit $K\rightarrow\infty$. From the self-consistent equations, when the coupling strength is sufficiently large if the system converges to the complete synchronization as $r\rightarrow1$ and all the oscillators are locked, we have $q\approxeq K \gg 1$. Then  Eq.~\eqref{eq_self_consistent3b} reads
\begin{equation}\label{eq_limit}
r\sin\alpha\approx\int_{-qb_{S,L}(a)+D\Omega^r}^{qb_{S,L}(a)+D\Omega^r}g_\Omega(\Omega) \frac{\Omega-D\Omega^r}{q}d\Omega
\end{equation}
where $qb_{S,L}(a)+D\Omega^r\gg0$ and $-qb_{S,L}(a)+D\Omega^r\ll0$ are the self-consistent conditions for the complete synchronization. From the property that $g_\Omega(\Omega)$ is a normalized distribution, we have the solution
\begin{equation}\label{eq_solution}
D\Omega_r\approx\bar{\Omega}-K\sin\alpha\approx\bar{\Omega}-q\sin\alpha,
\end{equation}
where $\bar{\Omega}$ is the mean-frequency of natural frequencies $\Omega$. From the symmetry of $g_\Omega(\Omega)$ as we assumed, we have $\bar{\Omega}=0$. Note that the collective frequency $\Omega_r$ depends on the coupling strength.

Such complete synchronization state only exists if the self-consistent conditions
\begin{equation}\label{Condition}
qb_{S,L}(a)+D\Omega^r\gg 0, \ -qb_{S,L}(a)+D\Omega^r\ll 0,
\end{equation} 
are satisfied. When $q$ is sufficiently large, we have $qb_L= q$ and $qb_S\approx C\sqrt{q}$ with the constant $C=4D/(\pi\sqrt{m})$. From the solution $D\Omega_r=-q\sin\alpha$, we deduce that only in the backward process with $qb_L=q$ the self-consistent conditions Eq.~\eqref{Condition} are satisfied if $\alpha\neq\pm\pi/2$ and hence the complete synchronization states exist. On the contrary, in the forward process, both expressions in Eq.~\eqref{Condition} are either positive or negative, depending on the value of $\alpha$. In both cases, the self-consistent conditions are not satisfied and oscillators cannot converge to the complete synchronization states. 

The critical coupling strength $K_n$ for this new synchronization process can be estimated by $qb_{S,L}(a)=|D\Omega^r|$, which gives
\begin{equation}
K_n=\frac{16D^2}{\pi^2m\sin^2\alpha}.
\end{equation}
When $\alpha\rightarrow0$ or $m\rightarrow0$, we have $K_n\rightarrow\infty$. In this case, all the oscillators are already synchronized with each other and therefore this new synchronization process does not manifest. The new synchronization process only appears in the forward process when $K_n$ is smaller than the critical point of the appearance of complete synchronization states. Hence one gets the usual synchronization processes with either small inertias or small phase shifts. 

In addition, when both $\alpha$ and $m$ are large enough, another effect of inertias should also be included, namely the appearance of additional synchronized clusters. As discussed in \cite{Olmi2014,Gao2018}, for second-order oscillators with large enough inertias, the steady states with only one cluster are not stable and several additional clusters can form besides the central cluster. In this case, the amplitude of the order parameter $r(t)$ exhibits a periodic oscillation. This kind of state is called \emph{oscillating state}, and is the direct result of inertias \cite{Gao2018}. Hence, as shown in Fig.~\ref{Fig_oscillating}, numerical simulations reveal that the synchronization process converges to oscillating states and not to the steady states calculated with the self-consistent method. In this case, the oscillators form two major synchronized clusters. We name this synchronization process the oscillating synchronization process to distinguish it from the classic synchronization process that leads to the complete synchronization state.

The numerical results for $N=10000$ oscillators are shown in Fig.~\ref{Fig_oscillating}. The natural frequencies of oscillators are chosen randomly from a Gaussian distribution. The inertia and phase shift of the oscillators are $m=2$, $\alpha=0.5$. Both the forward and backward processes are considered in the region $K\in[0,20]$ with $dK=0.1$. Comparing with the theoretical results from the self-consistent equations Eq.~\eqref{eq_self_consistent}, the numerical result in the backward process coincides well with the result from Eq.~\eqref{eq_self_consistent}, as shown in Fig.~\ref{Fig_oscillating}(a). However, in the forward process, the order parameter exhibits a large oscillation. For a specific state in the forward process at $K=16$, we show the order parameter $r(t)$, and mean-frequency $\bar\omega$ in Fig.~\ref{Fig_oscillating}(b,c). We see the periodic oscillation of $r(t)$ and correspondingly the multi-synchronization clusters shown as the stairs in Fig.~\ref{Fig_oscillating}(c). To check the properties of the oscillating state, we show the mean-frequency of the largest two clusters in the forward process. As shown in Fig.~\ref{Fig_oscillating}(d), the two mean-frequencies of these clusters depend linearly on the coupling strength $K$. Though the results in Fig.~\ref{Fig_oscillating} are shown up to $K=20$, we have checked that these non-synchronized oscillating states are still stable up to $K=500$.

Recall that when there is no phase shift, these synchronized clusters will merge into a single one for sufficiently large coupling strength \cite{Gao2018}. However, due to the phase shift $\alpha$ the separation of such clusters is strengthened. The frequency of these two clusters depends on the coupling strength $K$ approximately linearly with a slope proportional to the fraction of oscillators in it, as shown in Fig.~\ref{Fig_oscillating}(d). As a matter of fact, these two clusters cannot be synchronized by increasing the coupling strength. As a simple model exhibiting the same behaviour, consider a special system with only two values of natural frequencies, i.e. $N_1$ oscillators with $\Omega_1$ and $N_2$ oscillators with $\Omega_2$, following
\begin{equation}
  	m\ddot{\theta_i}+D\dot{\theta_i}=\Omega_1+\frac{K}{N}\sum_{j=1}^{N}\sin(\theta_j-\theta_i-\alpha),
\end{equation}
when $i=1,\dots,N_1$, and
\begin{equation}
  	m\ddot{\theta_i}+D\dot{\theta_i}=\Omega_2+\frac{K}{N}\sum_{j=1}^{N}\sin(\theta_j-\theta_i-\alpha), 
\end{equation}
when $i=N_1+1,\dots,N=N_1+N_2.$ The oscillators are naturally divided into two groups and synchronized within each group. The dimension of the system can be reduced and one finds 
\begin{subequations}
	\begin{align}
	&m\ddot{\theta_1}+D\dot{\theta_1}=\Omega_1-Kn_1\sin\alpha+Kn_2\sin(\theta_2-\theta_1-\alpha),\\
	&m\ddot{\theta_2}+D\dot{\theta_1}=\Omega_2-Kn_2\sin\alpha+Kn_1\sin(\theta_1-\theta_2-\alpha),
	\end{align}
\end{subequations}
where $\theta_1$ and  $\theta_2$ are the common phases of the oscillators in the first and second group respectively, and $n_1=N_1/N, n_2=N_2/N$ with $n_1+n_2=1$. With the definition of phase difference $\varphi=\theta_1-\theta_2$ we have
\begin{equation}\label{eq_two_1}
\begin{aligned}
	m\ddot{\varphi}+D\dot{\varphi}=&\Omega_1-\Omega_2-K(n_1-n_2)\sin\alpha\\
	&-K[n_2\sin(\varphi+\alpha)+n_1\sin(\varphi-\alpha)].
	\end{aligned}
\end{equation}
Without loss of generality, taking $n_1>n_2$, Eq.~\eqref{eq_two_1} can be rewritten as
\begin{equation}\label{eq_two_2}
m\ddot{\varphi}+D\dot{\varphi}=\Delta\Omega-\bar{K}\sin(\varphi+\bar{\alpha}),
\end{equation}
where
\begin{subequations}
	\begin{align}
	\Delta\Omega&=\Omega_1-\Omega_2-K(n_1-n_2)\sin\alpha,\\
	\bar{K}&=K\sqrt{\cos^2\alpha+(n_1-n_2)^2\sin^2\alpha}\equiv Kq(\alpha),\\
	\bar{\alpha}&=\arcsin\left(\frac{(n_1-n_2)\sin\alpha}{\sqrt{\cos^2\alpha+(n_1-n_2)^2\sin^2\alpha}}\right).
	\end{align}
\end{subequations}
The simplified dynamics in Eq.~\eqref{eq_two_2} is the same as the dynamics for second-order oscillators in the mean field Eq.~\eqref{eq_single_dynamic}. Hence the synchronization condition for the two clusters is determined by the two parameters
\begin{equation}\label{dynamics_2}
	a=\frac{D}{\sqrt{Kq(\alpha)m}},\ \ b=\frac{\Omega_1-\Omega_2}{Kq(\alpha)}-\frac{(n_1-n_2)\sin\alpha}{q(\alpha)}.
\end{equation}
As a result, with $K\rightarrow\infty$, we have $a\rightarrow0$ and $|b|\rightarrow(n_1-n_2)\sin\alpha/q(\alpha)>0$. From the fact that $b_S(a)\rightarrow0$ in the limit $a\rightarrow0$, the synchronization condition $|b|<b_S(a)$ can not be satisfied with increasing $K$. In this case, we have the non-synchronized process, where the two clusters cannot be synchronized. 

It is clear that the non-synchronized process is due to the cross-effect of inertia and phase shift. If $\alpha=0$, we have $q(\alpha)=1$. Substitution of $\alpha$ and $q(\alpha)$ into Eq.~\eqref{dynamics_2} yields 
\begin{equation}
a=\frac{D}{\sqrt{Km}},\ \ b=\frac{\Omega_1-\Omega_2}{K}.
\end{equation}
In the limit $K\rightarrow\infty$, we have $b_S\approx 4D/\sqrt{Km}\pi$. Hence no matter how large $|\Omega_1-\Omega_2|$ we always have $K_c=\pi^2(\Omega_1-\Omega_2)^2m/16D^2$ where the two clusters will become synchronized with $K>K_c$ with increasing $K$. On the other hand, if $m=0$ one gets $b_S=1$. From the fact that $(n_1-n_2)\sin\alpha/q(\alpha)<1$ we have $|b|<1=b_S$ in the limit $K\rightarrow\infty$. As a result, these two clusters will be synchronized when $K$ is large enough. 

In addition, similar to the analysis for $m=0$, in the backward processes with $b_L\equiv 1$, the synchronization states are not affected by the inertias and phase shifts. As a result, from the quite different properties of $b_L$ and $b_S(a)$, we have the non-trivial bi-stability of complete synchronization and oscillating states.

To test the conclusion above, we calculate the phase diagram of non-synchronized oscillating states. $N=10000$ oscillators are considered with a Gaussian distribution of their natural frequencies. With different inertias $m\in[0.1,4]$ and phase shifts $\alpha\in[0.01,1]$, we follow the oscillators in the forward process to a sufficient large coupling $K=40$. The boundary between oscillating states and partial synchronization states is determined by $\pm0.1$ standard deviations of the order parameter $r$ from its mean value. The result is shown in Fig.~\ref{Fig_oscillating00 }(a). The oscillating states exist when both the inertia and phase shift are relatively large. The fitting curve for the boundary lines read $\alpha=(\pi/2)/(1+2.96k)$. In Fig.~\ref{Fig_oscillating00 }(b), we check the basin of attraction of the oscillating state at $K=20$. The  oscillators sorted and separated into two groups according to their natural frequencies. The fraction of the two groups is defined by $n_1$ and $n_2$ with $n_1+n_2=1$. The initial frequencies of the oscillators are chosen randomly from a small region around $\Omega_1$ and $\Omega_2$, and their initial phases are chosen randomly from $[0,2\pi]$. Without loss of generality, we take $\Omega_1=-1$. From the numerical simulations, we see that there is clear large basin of attraction of the oscillating state as shown in Fig.~\ref{Fig_oscillating00 }(b). As we can see from the expression for the parameters $a,b$ in Eq.~\eqref{dynamics_2}, the basin of attraction of oscillating states is closely related to the frequency and fraction separation $|\Omega_1-\Omega_2|$ and $|n_1-n_2|$ of two groups.

\begin{figure}
	\centering
	\includegraphics[width=0.9\textwidth]{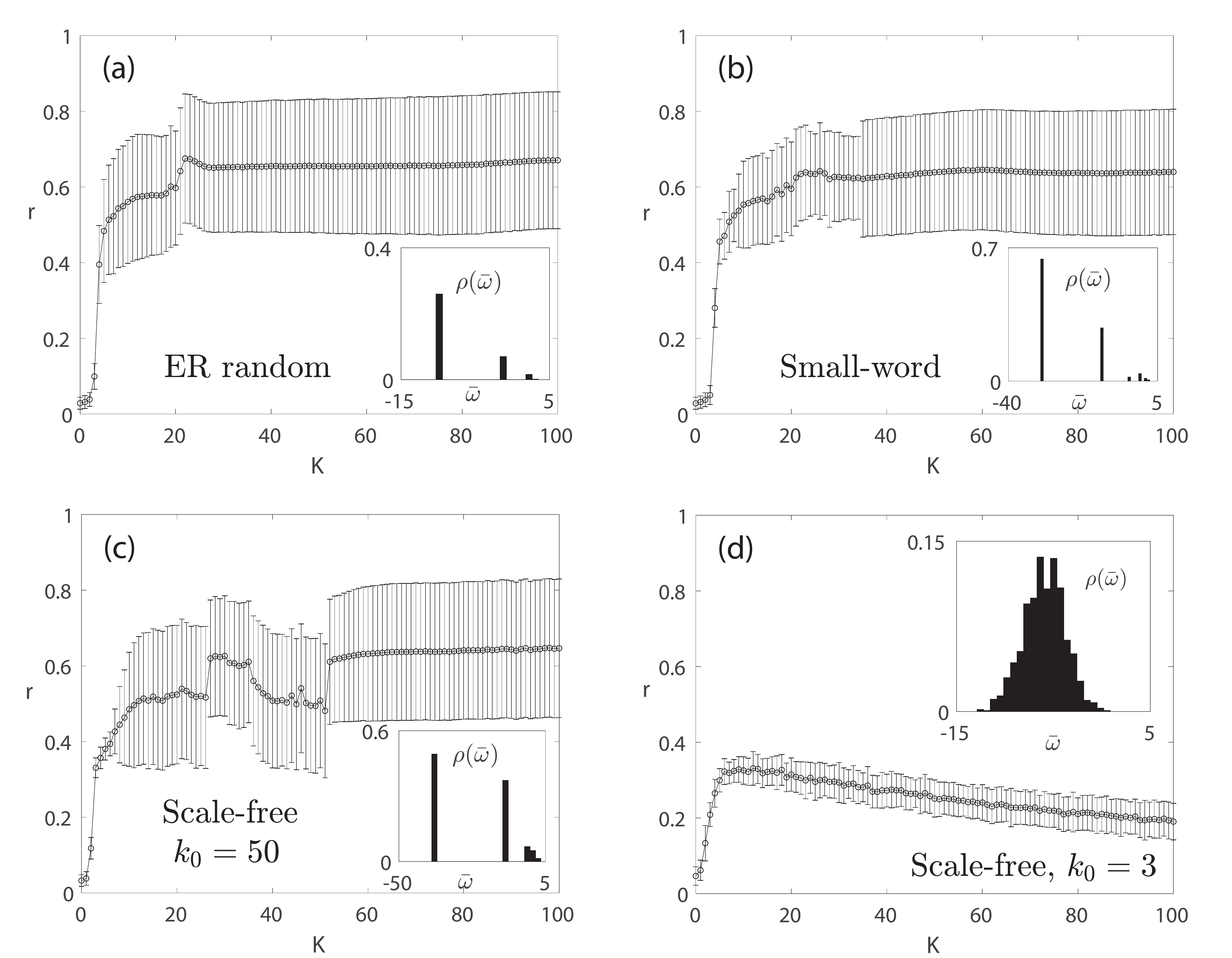}
	\caption{Synchronization transitions for $1000$ oscillators with a Gaussian distribution $g_\Omega(\Omega)=\frac{1}{\sqrt{2\pi}}e^{-\Omega^2/2}$  with $m=2, \alpha=0.5$ in backward process on (a) Erdos-R\'{e}nyi random networks \cite{Erdos1960evolution} with $p=0.3$ the probability for edge creation, (b) Watts-Strogatz small-world networks \cite{Watts1998collective} with $k=100$ the nearest connection in a ring and $p=0.3$ the probability for edge creation, (c) Barab\'{a}si-Albert scale-free networks \cite{Barabasi1999emergence} with the minimum degree $k_0=50$, (d) Barab\'{a}si-Albert scale-free networks with the minimum degree $k_0=3$.}\label{Fig_oscillating0}
\end{figure}

To check the generality of the oscillating synchronization process, we considered various systems of the second-order Kuramoto-Sakaguchi oscillators. For all-connected oscillators, this new oscillating synchronization process is found in all the cases we considered, including uniform, Lorentz and double-Gaussian/Lorentz distributions of the natural frequencies. For oscillators in complex networks, we consider scale-free, ER random, and small-word networks. The non-synchronization processes are found in all the systems as shown in Fig.~\ref{Fig_oscillating0}. The oscillating states appear in these processes when the mean-degree of such networks is large. On the other hand, with a smaller mean-degree, the second synchronization cluster is suppressed by the topology of the network. We find the non-synchronized steady states converge to $r=0$ in the limit $K\rightarrow\infty$ as shown in Fig.~\ref{Fig_oscillating0}(d). The oscillating synchronization process depends on the mean-degree of networks, not their densities. This fact is closely related to the conditions determining weather the mean-field assumption works for random networks. The suppression of the oscillating states is beyond the scope of this paper, and will be considered in a forthcoming work.

\section{Conclusion}\label{section_four}

In this paper we analyse the second-order oscillators with phase shifts, namely second-order Kuramoto-Sakaguchi model. The self-consistent method is generalized and used to study the steady states of oscillators. With the inertia introduced phase shifts, the non-universal transitions of Kuramoto-Sakaquchi oscillators \cite{Omel2012} are canceled out by a small value of $m$. The changing of abrupt to continuous transitions with the effect of inertias proposed in \cite{Barre2016} is also shown and studied by the self-consistent method.

In addition, the cross-effect of inertia and phase shifts also results in the oscillating synchronization forward processes. Instead of synchronization states, the system will stay in the oscillating state and can not be synchronized with increasing coupling strength. This interesting phenomenon is due to the combination of additional synchronized clusters as an effect of inertias and the dependence of $\Omega^r$ on $K$ as an effect of phase shifts. Using numerical simulations, such non-synchronized processes are also found in different distributions of natural frequencies and topologies of the networks.

\begin{acknowledgments}
We would like to thank the Center for Information Technology of the University of
Groningen for their support and for providing access to the Peregrine high
performance computing cluster. J. Gao would like to acknowledge scholarship support from the China Scholarship Council (CSC).
\end{acknowledgments}

\bibliographystyle{elsarticle-num-names}
\bibliography{papersnew}

\end{document}